\documentclass[10pt,aps,twocolumn,eqsecnum,nofootinbib,floatfix,amsmath,amssymb]{revtex4}
\usepackage{graphicx}
\usepackage{dcolumn}
\usepackage{bm}
\usepackage{epsfig,psfrag,amsmath,amssymb,float}
\input{epsf}
\begin{document}

\def\eg{{\it e.g.~}}
                                                                                
\title{THE QUANTUM MECHANICS OF CLOSED SYSTEMS\footnote{A slightly
shortened version of an article that appeared 
in the festschrift for C.W.~Misner,  ed.~by B.L.~Hu, M.P.~Ryan and
C.V.~Vishveshwara, Cambridge University Press, Cambridge (1993).}}
                                                                                
\author{James B. Hartle}
\email{hartle@physics.ucsb.edu}
\affiliation{Department of Physics, University of California\\
 Santa Barbara, CA 93106-9530 USA}
                                                                                
                                                                                
\begin{abstract}

A pedagogical introduction is given to the quantum mechanics of closed
systems, most generally the universe as a whole.  Quantum mechanics aims
at predicting the probabilities of alternative coarse-grained time
histories of a closed system.  Not every set of alternative
coarse-grained histories that can be described may be consistently
assigned probabilities because of quantum mechanical interference
between individual histories of the set.  In ``Copenhagen'' quantum
mechanics, probabilities can be assigned to histories of a
subsystem that have been ``measured''.  In the quantum mechanics of
closed systems, containing both observer and observed,
probabilities are assigned to those sets of alternative histories for
which there is negligible interference between individual histories 
 as a consequence of the system's initial condition and dynamics.
Such sets of histories are said to decohere.  We define decoherence for
closed systems in the simplified case when quantum gravity can be
neglected and the initial state is pure.  Typical mechanisms of
decoherence that are widespread in our universe are illustrated.

Copenhagen quantum mechanics is an approximation to the more general
quantum framework of closed subsystems.  It is appropriate when there is
an approximately isolated subsystem that is a participant in a
measurement situation in which (among other things) the decoherence of
alternative registrations of the apparatus can be idealized as exact.

Since the quantum mechanics of closed systems does not posit the
existence of the quasiclassical realm of everyday experience, the
domain of the approximate aplicability of classical physics must be
explained.  We describe how a quasiclassical realm described by
averages of densities of approximately conserved quantities could be an
emergent feature of an initial condition of the universe that implies
the approximate classical behavior of spacetime on accessible scales.

\end{abstract}
                                                                                
\maketitle

\section{Introduction}

It is an inescapable inference from the physics of the last sixty years
that we live in a quantum mechanical universe --- a world in which the basic
laws of physics conform to that framework for prediction we call quantum
mechanics. We perhaps have little evidence of peculiarly quantum
mechanical phenomena on large and even familiar scales, but there is no
evidence that the phenomena that we do see cannot be described in
quantum mechanical terms and explained by quantum mechanical laws.
If this inference is correct, then there must be a description of the universe
as a whole and everything in it in quantum mechanical terms.  The nature
of this description and its observable consequences are the subject of
quantum cosmology

Our observations of the present universe on the largest scales are
crude and a
classical description of them is entirely adequate.  Providing a quantum
mechanical description of these observations alone might be an
interesting
intellectual challenge, but it would be unlikely to yield testable
predictions
differing from those of classical physics.  Today, however, we have a
more
ambitious aim.  We aim, in quantum cosmology, to provide a theory of the
initial condition of the universe which will predict testable
correlations
among observations today.  There are no realistic predictions of any
kind that
do not depend on this initial condition, if only very weakly.
Predictions of
certain observations may be testably sensitive to its details.  These
include
the large scale homogeneity and isotropy of the universe, its
approximate
spatial flatness, the spectrum of density fluctuations that produced the
galaxies, the homogeneity of the thermodynamic arrow of time, and the
existence
of classical
spacetime.  Recently, there has been speculation that even
the coupling constants of the effective interactions of the
elementary
particles at accessible energy scales may be probabilistically
distributed with a distribution which may  depend, in part, on the
initial
condition of the universe \cite{Haw83, Col88, GS88}.
  It is for such reasons that the search for a
theory
of the initial condition of the universe is just as necessary and just
as
fundamental as the search for a theory of the dynamics of the elementary
particles. They may even be the same searches.

The physics of the very early universe is likely to be quantum
mechanical in an
essential way.  The singularity theorems of classical general
relativity
suggest that an early era preceded ours in which even the geometry of
spacetime
exhibited significant
quantum fluctuations.  It is for a theory of the initial condition
that describes this era, and all later ones, that we need to spell out
how to apply  quantum
mechanics to cosmology. Recent years have seen much promising progress
in the search for a theory of the quantum initial condition.  However,
it is not my purpose to review these developments here.\footnote{For
a recent review of quantum cosmology see \cite{Hal91}.}
Rather, I shall
argue that this somewhat obscure branch of astrophysics may
have implications for the formulation and interpretation of quantum
mechanics on  day-to-day scales.  My thesis will be that by looking at
the universe as a whole one is led to an understanding of quantum
mechanics which clarifies many of the long standing interpretative
difficulties of the subject.

The Copenhagen frameworks for quantum mechanics, as they were formulated
in the '30s and '40s and as they exist in most textbooks today, are
inadequate for quantum cosmology.
 Characteristically these formulations
assumed, as {\it external} to the framework of wave function and
Schr\"odinger equation, the quasiclassical realm we see all about us.  Bohr
\cite{Boh58}  
spoke of phenomena which could be alternatively described in classical
language.  In their classic text, Landau and Lifschitz \cite{LL58}
 formulated
quantum mechanics in terms of a separate classical physics.  Heisenberg
and others stressed the central role of an external, essentially
classical, observer.\footnote{For a clear statement of
this point of view, see \cite{LB39}.}
Characteristically, these formulations assumed a
possible division of the
world into
``obsever'' and ``observed'', assumed that ``measurements'' are the
primary
focus of
scientific statements and, in effect,  posited the existence of an
external
``quasiclassical realm''.  However, in a theory of the whole thing there can
be no
fundamental division into observer and observed.  Measurements and
observers
cannot be fundamental notions in a theory that seeks to describe the
early
universe when neither existed.  In a basic formulation of quantum
mechanics
there is no reason in general for there to be any variables that exhibit
classical behavior in all circumstances.
  Copenhagen quantum mechanics thus needs to be generalized to
provide a quantum framework for cosmology.

In a generalization of quantum mechanics which does not {\it posit} the
existence of a quasiclassical realm, the domain of
 applicability of classical physics
must be {\it explained}.
For a quantum mechanical system to exhibit classical behavior there must
be
some restriction on its state and some coarseness in how it is
described.
This is clearly illustrated in the quantum mechanics of a single
particle.  Ehrenfest's theorem shows that generally
\begin{equation}
M\frac{d^2\langle x\rangle}{dt^2} = \left \langle -\frac{\partial
V}{\partial
x}\right \rangle.
\label{oneone}
\end{equation}
However, only for special states, typically narrow wave packets, will
this become an equation of motion for $\langle x \rangle$ of the form
\begin{equation}
M\frac{d^2\langle x \rangle}{dt^2} = -\frac{\partial V(\langle x
\rangle)}{\partial x}.
\label{onetwo}
\end{equation}
For such special states, successive observations of position in time
will exhibit the classical correlations predicted by the equation
of motion \eqref{onetwo} {\it provided} that
these observations are coarse enough so that the properties of the state
which allow \eqref{onetwo} to replace  the general relation
\eqref{oneone} are not affected by these observations.  An {\it exact}
determination of position, for example,  would yield a completely
delocalized wave packet an
instant later and \eqref{onetwo} would no longer be a good approximation to
\eqref{oneone}.  Thus, even for large systems, and in particular for the universe
as a whole, we can expect classical behavior only for certain initial states
and then only when a sufficiently coarse grained description is used.

If classical behavior is {\it in general}  a consequence only of a certain
class of states  in quantum
mechanics, then, as a particular case, we can expect to have classical
spacetime
only for certain states in quantum gravity.  The classical spacetime
geometry
we see all about us in the late universe is not property of every state
in a
theory where geometry fluctuates quantum mechanically. Rather, it is
traceable
fundamentally to restrictions on the initial condition.
Such restrictions are likely to be generous in that, as in the single
particle
case, many different states will exhibit classical features. The
existence
of classical spacetime and the applicability of classical physics
 are thus not likely to be very restrictive conditions on constructing
a theory of the initial condition. 

It was Everett who, in 1957, first suggested how to generalize the
Copenhagen frameworks so as to apply quantum mechanics to 
cosmology.\footnote{The original reference is \cite{Eve57}. For
a useful collection of reprints see \cite{DG73}.}
Everett's idea was to take quantum mechanics seriously and apply it
to the
universe as a whole.  He showed how an observer could be considered part
of
this system and how its activities --- measuring, recording, calculating
probabilities, etc. --- could be described within quantum mechanics.
Yet the
Everett analysis was not complete.  It did not adequately describe
within
 quantum mechanics the origin of the ``quasiclassical realm'' of familiar
experience\footnote{In our earlier work \eg \cite{GH90a} this was called
a ``quasiclassical domain'', but this risked confusion with the use of
the word ``domain'' in condensed matter physics.} 
 nor, in an observer independent way, the meaning of the
``branching''
that replaced the notion of measurement.  It did not distinguish from
among
the vast number of choices of quantum mechanical observables that are in
principle available to an observer, the particular choices that, in
fact,
describe the
quasiclassical relm.

 In this essay, I will describe joint work with Murray Gell-Mann \cite{GH90a,
GH90b}
which aims at a coherent formulation of quantum mechanics for the
universe as a whole that is a framework to explain rather than  posit the
quasiclassical realm of everyday experience.  It is an attempt at an
extension, clarification, and completion of the Everett interpretation.
It builds on many aspects of the, so called post-Everett development,
especially the work of Zeh \cite{Zeh71}, Zurek \cite{Zur81, Zur82}, and
 Joos and Zeh \cite{JZ85}.  At important
points it coincides with the, independent, earlier work of Bob 
Griffiths \cite{Gri84}
and Roland Omn\`es (\eg as reviewed in  
\cite{Omn92}).  

Our work is not complete, but I hope to sketch how
it might become so.  It is by now a very long story but I will try 
to describe
the important parts in simplified terms.

\section{Probabilities in General and Probabilities in Quantum
Mechanics}

Even apart from quantum mechanics, there is no certainty in this
world and
therefore physics deals in
probabilities.
It deals most generally with the probabilities for
alternative
time histories of the universe.  From these, conditional probabilities
can be constructed that are 
appropriate when some features about our specific history are known 
and further ones are to be predicted.

 To understand what probabilities mean for a single closed system, 
it is best to understand
how they
are used.  We deal, first of all, with probabilities for {\it single}
events
of the {\it single} system.
When these
probabilities become sufficiently close to zero or one there is a
definite
prediction on which we may act.  How sufficiently close to 0 or 1 the
probabilities must be depends on the circumstances in which they are
applied.
There is no certainty that the sun will come up tomorrow at the time
printed in
our daily newspapers.  The sun may be destroyed by a neutron star now
racing
across the galaxy at near light speed.  The earth's rotation rate could
undergo
a quantum fluctuation.  An error could have been made in the computer
that
extrapolates the motion of the earth.  The printer could have made a
mistake in
setting the type.  Our eyes may deceive us in reading the time.  Yet, we
watch the sunrise at the appointed time because we compute, however
imperfectly, that the probability of these things happening is
sufficiently
low.

 Various strategies can be employed to identify situations where
probabilities
are near zero or one.  Acquiring information and considering the
conditional
probabilities based on it is one such strategy.  Current theories of the
initial condition of the
universe predict almost no probabilities near zero or one without
further
conditions.  The ``no
boundary'' wave function of the universe, for example, does not predict
the
present position of the sun on the sky.  However, it will predict  that
the
conditional probability for the sun to be at the position predicted by
classical celestial mechanics given a few previous positions is a number
very
near unity.

Another strategy to isolate probabilities near 0 or 1 is to consider
ensembles
of repeated observations of identical subsystems in the closed system.  
There are no
genuinely
infinite ensembles in the world so we are necessarily concerned with the
probabilities for deviations of the behavior of a finite ensemble
from the expected behavior of an infinite one.  These are probabilities
for a
single feature (the deviation) of a single system (the whole ensemble).

 The existence of large ensembles of repeated observations in identical
circumstances and their ubiquity in laboratory science should not,
therefore,
obscure the
fact that in the last analysis physics must predict probabilities for
the
single system that is the ensemble as a whole.  Whether it is the
probability
of a successful marriage, the probability of the present galaxy-galaxy
correlation function, or the probability of the fluctuations in an
ensemble of
repeated observations, we must deal with the probabilities of single
events in
single systems.
In geology, astronomy, history, and cosmology, most predictions of
interest
have this character.
The  goal of physical theory is, therefore, most generally to predict
the
probabilities of histories of single events of a single system.

 Probabilities need be assigned to histories by physical theory only
up to the
accuracy they are used.  Two theories that predict probabilities for the
sun
not rising tomorrow at its classically calculated time that are both
well
beneath the standard on which we act are equivalent for all practical
purposes
as far as this prediction is concerned. It is often convenient,
therefore, to deal with approximate probabilities which satisfy the
rules of probability theory up to the standard they are used.

The characteristic feature of a quantum mechanical theory is that not
every set of alternative
histories that may be described can be assigned  probabilities.  Nowhere is
this
more clearly illustrated than in the two slit experiment illustrated in
Figure 1.  In the usual
``Copenhagen'' discussion if we have 
 not measured which of the two slits
the
electron passed through on its way to being detected at the screen, then
we are not permitted to assign probabilities to these alternative
histories.
It would be inconsistent to do so since the correct probability sum rule
would not be satisfied.  Because of interference, the probability to
arrive at point $y$ on the screen is not the sum of the probabilities to arrive at $y$ going
through
the upper or lower slit:
\begin{equation}
p(y) \not= p_U (y) + p_L (y) 
\label{twoone}
\end{equation}
because in quantum theory probabilities are squares of amplitudes and 
\begin{equation}
|\psi_L (y) + \psi_U (y) |^2 \not= |\psi_L (y) |^2 + |\psi_U (y) |^2\, .
\label{twotwo}
\end{equation}

\begin{figure}[t]
\begin{center}
\epsfig{file=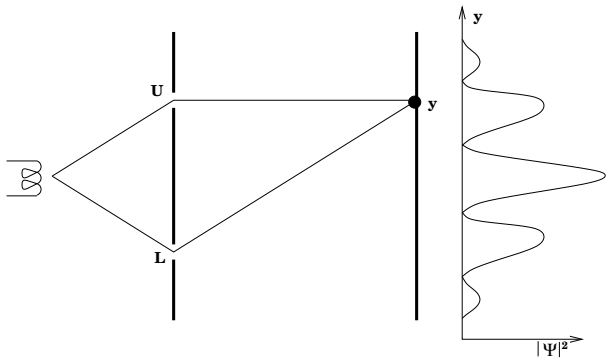,width=8cm,clip}
\caption{The two-slit experiment.  An
electron gun at left emits an electron traveling towards a screen with
two slits, its progress in space recapitulating its evolution in time.  When
precise detections are made of an ensemble of such electrons at the
screen it
is not possible, because of interference, to assign a probability to the
alternatives of whether an individual electron went through the upper
slit or
the lower slit.  However, if the electron interacts with apparatus that
measures which slit it passed through, then these alternatives decohere
and
probabilities can be assigned.}
\end{center}
\end{figure}
If we {\it have} measured which slit the electron went through, then the
interference is destroyed, the sum rule obeyed, and we {\it can}
meaningfully
assign probabilities to these alternative histories.

A rule is thus needed in quantum theory to determine which sets of
alternative histories may be assigned probabilities and which may not.
In Copenhagen quantum mechanics, the rule is that probabilities are
assigned to histories of alternatives of a subsystem 
that are {\it measured} and not in
general otherwise.

\section{Probabilities for a Time Sequence of Measurements}

To establish some notation, let us review in more detail the usual
rules for the probabilities of time sequences of ideal measurements
of subsystem using the two-slit experiment of Figure 1
 as an example.

Alternatives for the electron
 are represented by projection operators in its Hilbert space. 
 Thus, in the two
slit experiment, the alternative that the electron passed through the
lower slit is represented by the projection operator
\begin{equation}
P_U = \Sigma_s \int_U d^3x \left|\vec x, s\rangle\langle\vec x, 
s\right|
\label{threeone}
\end{equation}
where $|\vec x, s\rangle$ is a localized state of the electron with
spin component $s$, and the integral is over a volume around the upper
slit.  There is a similar projection operator $P_L$ for the alternative
that the electron goes through the lower slit.  These are exclusive
alternatives and they are exhaustive.  These properties, as well as the
requirements of being projections, are represented by the
relations
\begin{equation}
P_L P_U = 0\ , \quad P_U + P_L = 1,
\quad P_L^2 = P_L, \quad P^2_U = P_U \ .
\label{threetwo}
\end{equation}
There is a similarly defined set of projection operators $\{P_y\}$
representing the alternative positions of arrival at the screen.

We can now state the rule for the joint probability that the electron
initially in a state $|\psi(t_0)\rangle$ at $t=t_0$ is determined by an
ideal measurement at time $t_1$ to have passed through the upper slit and
measured at time $t_2$ to arrive at point $y$ on the screen.  If one
likes, one can imagine the case in which the electron is in a 
narrow wave packet
in the horizontal direction with a velocity defined as sharply as possible
consistent with the uncertainty principle.  The joint probability is
negligible unless $t_1$ and $t_2$ correspond to the times of flight to
the slits and to the screen respectively.

The first step in calculating the joint probability is to evolve the state of
the electron to the time $t_1$ of the first measurement
\begin{equation}
\bigl | \psi (t_1)\bigr\rangle = e^{-iH(t_1-t_0)/\hbar}\bigr|
\psi(t_0)\bigr\rangle\, . 
\label{threethree}
\end{equation}
The probability that the outcome of the measurement at time $t_1$ 
is that the electron passed through the upper slit is:
\begin{equation}
({\rm Probability\ of}\ U) = \left\Vert P_U\big |\psi(t_1)\big\rangle
\right\Vert^2 
\label{threefour}
\end{equation}
where $\Vert \cdot\Vert$ denotes the norm of a vector in the electron's
Hilbert space.
If the outcome was the upper slit, and the measurement was an ``ideal''
one, that disturbed the electron as little as possible in making its
determination, then after the measurement the state vector is reduced to
\begin{equation}
\frac{P_U |\psi(t_1)\rangle}{\Vert P_U |\psi(t_1)\rangle\Vert}\ . 
\label{threefive}
\end{equation}
This is evolved to the time of the next measurement
\begin{equation}
|\psi(t_2)\rangle =
e^{-iH(t_2-t_1)/\hbar}\frac{P_U|\psi(t_1)\rangle}{\Vert P_U
|\psi(t_1)\rangle\Vert}\, . 
\label{threesix}
\end{equation}
The probability of being detected at point $y$ on the screen at time
$t_2$ {\it given} that the electron passed through the upper slit is
\begin{equation}
({\rm Probability\ of}\ y\ given\ U) = \left\Vert P_y
|\psi(t_2)\rangle\right\Vert^2\, .
\label{threeseven}
\end{equation}

The {\it joint} probability that the electron is measured to have gone
through
the upper slit {\it and} is detected at $y$ is the product of the
conditional probability \eqref{threeseven} with the probability \eqref{threefour}
that the electron passed through $U$.  The latter factor
cancels the denominator in
\eqref{threesix} so that combining all of the above equations in this
section, we have
\begin{multline}
({\rm Probability\  of}\ y\ and\ U) \\ 
= \left\Vert P_y
e^{-iH(t_2-t_1)/\hbar} P_U e^{-iH(t_1-t_0)/\hbar}\big |\psi(t_0)\big
\rangle\right\Vert^2\, . 
\label{threeeight}
\end{multline}
With Heisenberg picture projections this takes the even simpler
form
\begin{equation}
({\rm Probability\ of}  \ y\ {\rm and}\ U)  
=\left\Vert P_y (t_2) P_U (t_1)
\ \big |\psi(t_0)\rangle\right\Vert^2\, . 
\label{threenine}
\end{equation}
where, for example,
\begin{equation}
P_U(t) = e^{iHt/\hbar} P_U e^{-iHt/\hbar}\, . 
\label{threeten}
\end{equation}
The formula \eqref{threenine} is a compact and unified expression of the two
laws of evolution that characterize the quantum mechanics of measured
subsystems --- unitary evolution in between measurements and reduction
of the wave packet at a measurement.\footnote{As has been noted by
many authors, \eg \cite{Gro52} and \cite{Wig63} among the
earliest.}
  The important thing to remember
about the expression \eqref{threenine} is that everything in it ---
projections, state vectors, Hamiltonian --- refer to the Hilbert space
of a subsystem, in this example the Hilbert space of the 
electron that is measured.  

In ``Copenhagen'' quantum mechanics, it is
measurement that determines which histories of a subsystem can be assigned
probabilities and formulae like \eqref{threenine} that determine what these
probabilities are.  We cannot have such  rules in the quantum mechanics
of closed systems.
There is no fundamental division of a closed system into measured
subsystem and measuring apparatus.  There is no fundamental reason for
the closed system to contain classically behaving measuring apparatus in all
circumstances.  In particular, in the early universe none of these
concepts seem relevant.  We need a more observer-independent,
measurement-independent, quasiclassical realm-independent rule for which
histories of a closed system can be assigned probabilities and what
these probabilities are.  The next section describes this rule.

\section{Post-Everett Quantum Mechanics}

To describe the rules of post-Everett quantum mechanics, I shall make a
simplifying assumption.  I shall neglect gross quantum fluctuations in
the geometry of spacetime, and assume a fixed background spacetime geometry
which
supplies a definite meaning to the notion of time.  This  is
an excellent approximation on accessible scales for times later than
$10^{-43}$ sec after the big bang.  The familiar apparatus of Hilbert
space, states, Hamiltonian, and other operators may then be applied to
process of prediction.  Indeed, in this context the quantum mechanics of
cosmology is in no way distinguished from the quantum mechanics of a
large isolated box, perhaps expanding, but containing both the observed
and its observers (if any).  

\begin{figure}[t!]
\begin{center}
\epsfig{file=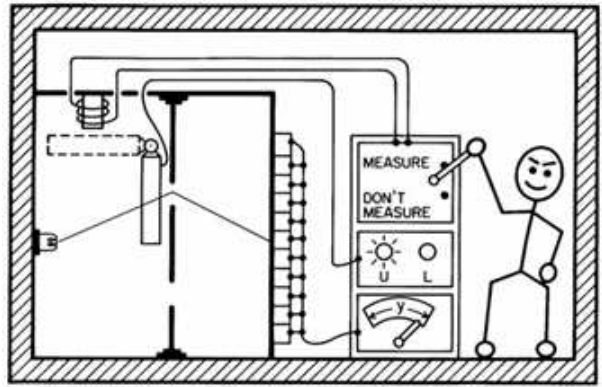,width=8cm,clip}
\caption{A model closed quantum system containing an observer together
with the necessary apparatus for carrying out a two-slit experiment.
Alternatives for the system include 
whether the observer measured which slit the electron passed through or
did not, whether the electron passed through the upper or lower slit,
the alternative positions of arrival of the electron at the screen, the
alternative arrival positions registered by the apparatus, the
registration of these in the brain of the observer, etc., etc., etc.  Each
exhaustive set of exclusive alternatives is represented by an exhaustive
set of orthogonal projection operators on the Hilbert space of the
closed system.  Time sequences of such sets of alternatives describe
sets of alternative coarse-grained histories of the closed system.
Quantum theory assigns probabilities to the individual alternative
histories in such a set when there is negligible quantum mechanical
interference between them, that is, when the set of histories
decoheres.
A more refined model might consider a quantity of matter in a closed
box.  One could then consider alternatives such as whether the box
contains a two-slit experiment or does not as well as alternative
configureatins of the atoms in the box.}
\label{box}
\end{center}
\end{figure}

A set of alternative histories for a closed system is specified by
giving exhaustive sets of exclusive alternatives at a sequence of times.
Consider a model closed system initially in a
pure state that can be described as an observer and two slit experiment,
with appropriate apparatus for producing the electrons, detecting which
slit they passed through, and measuring their position of arrival on the
screen (Figure 2).  Some alternatives for the whole system are:

\begin{enumerate}
\item Whether or not the observer decided to measure which slit the
electron went through.
\item Whether the electron went through the upper or lower slit.
\item The alternative positions, $y_1, \cdots, y_N$, that the
electron could have arrived at the screen.
\end{enumerate}
This set of alternatives at a sequence of times defines a set of
histories whose characteristic branching structure is shown in Figure 3.
An individual history in the set is specified by some particular
sequence of alternatives, \eg (measured, upper, $y_9$).

\begin{figure}[t!]
\begin{center}
\epsfig{file=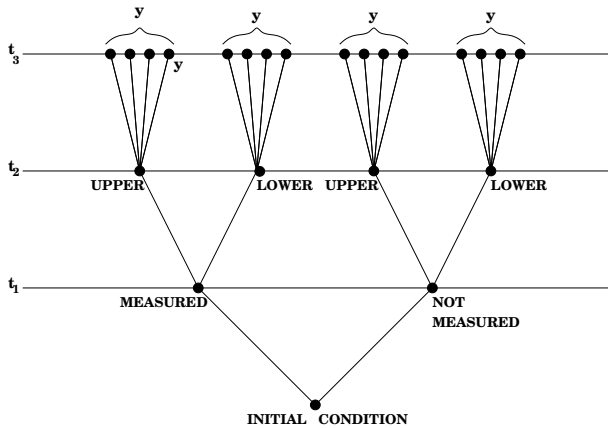,width=8cm,clip}
\caption{Branching structure of a set of
alternative histories.  This figure illustrates the set of alternative
histories of the closed system illustrated in Figure \ref{box} 
defined by the alternatives of whether the observer decided to
measure or did not decide to measure which slit the electron went
through at time $t_1$, whether the electron went through the upper slit
or through the lower slit at time $t_2$, and the alternative positions
of arrival $y$ at the screen at time  $t_3$.  A single branch corresponding to the
alternatives that the measurement was carried out, the electron went
through the upper slit, and arrived at point $y_9$ on the screen is
illustrated by the heavy line. \\ 
The illustrated set of histories does {\it not}
decohere because there is significant quantum mechanical interference
between the branch where no measurement was carried out and the electron
went through the upper slit and the similar branch where it went through
the lower slit.
A related set of histories that does decohere can be obtained by
replacing the alternatives at time $t_2$ by the following set of three
alternatives: (a record of the decision shows a measurement was
initiated and the electron went through
the upper slit); (a record of the decision shows a measurement was
initiated and the electron went through the lower slit); (a record of
the decision shows that the measurement was not initiated).  The
vanishing of the interference between the alternative values of the
record and the alternative configurations of apparatus ensures the
decoherence of this set of alternative histories.}
\end{center}
\end{figure}

Many other sets of alternative histories are possible for the closed
system.  For example, we could have included alternatives describing the
readouts of the apparatus that detects the position that the electron
arrived on the screen.  If the initial condition corresponded to a good
experiment there should be a high correlation between these alternatives
and the position that the electron arrives at the screen.  In a more
refined model we could discuss
alternatives corresponding to thoughts in the observer's brain, or to
the individual positions of the atoms in the apparatus, or to the
possibilities that these atoms reassemble in some completely different
configuration. There are a vast number of possibilities.

Characteristically the alternatives that are of use to us as observers
are very coarse grained,
distinguishing only very few of the degrees of freedom of a large closed
system.  This is especially true if we recall that our box with observer
and two-slit experiment is only an idealized model.  The most general
closed system is the universe itself, and, as I hope to show, the only
realistic closed systems are of cosmological dimensions.  Certainly, as
observers of the universe we
utilize only very, very coarse-grained descriptions of the universe as a
whole. 

I would now like to state the rules that determine which coarse-grained
sets of histories may be assigned probabilities and what those
probabilities are.  The essence of the rules I shall describe can be
found in the work of Bob Griffiths \cite{Gri84}.  
The general framework was extended
by Roland Omn\`es \cite{Omn92}
and was independently, but later, arrived at by Murray
Gell-Mann and myself \cite{GH90a}.  
The idea is simple: The failure of probability
sum rules due to quantum interference is the obstacle to assigning
probabilities.  {\it Probabilities can be assigned to just those sets of
alternative histories of a closed system for which there is negligible
interference between the individual histories in the set as a
consequence of the Hamiltonian and {\it particular} initial state 
the closed system has.  All probability 
sum rules {\it are} satisfied as a result of the absence of
interference.}  Let us now give this idea a precise expression.

Sets of alternatives at one moment of time are represented by sets of
orthogonal projection operators.  Employing the Heisenberg picture these
can be denoted $\{P^k_{\alpha_k} (t_k)\}$.  The superscript $k$ denotes
the set of alternatives being considered at time $t_k$ 
(for example, the set of alternative position
intervals $\{y_1, \cdots, y_N\}$ at which the electron might
 arrive at the screen at time $t_3$),  
$\alpha_k$ denotes the
particular alternative in the set (for example $y_9$)
 and $t_k$ is the time.  
The set of $P$'s satisfy
\begin{equation}
\sum\nolimits_{\alpha_k} P^k_{\alpha_k} (t_k) = 1\ ,\quad P^k_{\alpha_k}
(t_k) P^k_{\alpha^\prime_k} (t_k) = \delta_{\alpha_k\alpha^\prime_k}
P^k_{\alpha_k} (t_k)
\label{fourone}
\end{equation}
showing that they represent an exhaustive set of exclusive
alternatives.

Sets of alternative histories are defined by giving sequences of sets of
alternatives at definite moments of time, \eg
$\{P^1_{\alpha_1} (t_1)\}\ , \{P^2_{\alpha_2} (t_2)\}, \cdots,
\{P^n_{\alpha_n} (t_n)\}$. Different choices for $\{P^1_{\alpha_1}
(t_1)\}$, $\{P^2_{\alpha_2} (t_2)\}$, 
 etc.~describe different sets of alternative histories of the
closed system.  An individual history in a given set corresponds to a
particular sequence  $(\alpha_1, \cdots, \alpha_n)\equiv \alpha$ and,
for each history, there is a corresponding chain of projection
operators
\begin{equation}
C_\alpha \equiv P^n_{\alpha_n} (t_n) \cdots P^1_{\alpha_1} (t_1)
\, .
\label{fourtwo}
\end{equation}
For example, in the two slit experiment in a box illustrated in Figure 2,
the history in which the observer decided at time $t_1$ to measure which
slit the electron goes through, in which the electron goes through the
upper slit at time $t_2$, and arrives at the screen in position interval
$y_9$ at time $t_3$, would be represented by the chain
\begin{equation}
P^3_{y_9} (t_3) P^2_U (t_2) P^1_{\rm meas} (t_1)
\label{fourthree}
\end{equation}
in an obvious notation. The only difference between this situation and
that of the ``Copenhagen'' quantum mechanics of measured subsystems is
the following:  The sets of operators $\{P^k_{\alpha_k} (t_k)\}$
defining alternatives for the closed system act on the Hilbert space of
the closed system that includes the variables describing any apparatus,
observers, and anything else.  The operators defining alternatives in
Copenhagen quantum mechanics act only on the Hilbert space of the
measured subsystem.

When the initial state is pure, it can be resolved into {\it branches}
corresponding to the individual members of any set of alternative
histories.  The generalization to an impure initial density matrix is
not difficult \cite{GH90a}, but for simplicity we shall assume a pure initial
state throughout this article.  Denote the initial state by
$|\Psi\rangle$ in the Heisenberg picture.  Then
\begin{equation}
|\Psi\rangle = \sum\nolimits_\alpha C_\alpha |\Psi\rangle =
\sum\limits_{\alpha_1,\cdots, \alpha_n} P^n_{\alpha_n} (t_n) \cdots
P^1_{\alpha_1} (t_1) |\Psi\rangle\, . 
\label{fourfour}
\end{equation}
This identity follows by applying the first of \eqref{fourone} to all the
sums over $\alpha_k$ in turn.  The vector
\begin{equation}
C_\alpha|\Psi\rangle
\label{fourfive}
\end{equation}
is the {\it branch}\footnote{More specifically a {\it branch state
vector}.}  corresponding to the individual history $\alpha$ and
\eqref{fourfour} is the resolution of the initial state into branches.

When the branches corresponding to a set of alternative histories are
sufficiently orthogonal the set of histories is said to {\it decohere}.
More precisely a set of histories decoheres when
\begin{equation}
\langle \Psi | C^\dagger_{\alpha^\prime} C_\alpha |\Psi \rangle \approx
0\ ,\quad {\rm for\ \ any}\ \ \alpha^\prime_k \not= \alpha_k\, .
\label{foursix}
\end{equation}
We shall return to the standard with which decoherence should be
enforced, but first let us examine its meaning and consequences.

Decoherence means the absence of quantum mechanical interference between
the individual histories of a coarse-grained set.\footnote{The
term ``decoherence'' is used in several different ways in the literature.
Therefore, for those familiar with other work,
 a comment is in order to specify how we are employing the term in this
simplified presentation.  We have followed our previous work
\cite{GH90a}, \cite{GH90b} in using the term ``decoherence'' to refer
to a property of a set of alternative time {\it histories} of a closed
system.  A decoherent set of histories is one for which the quantum
mechanical interference between individual histories is small enough to
guarantee an appropriate set of probability sum rules. Different notions
of decoherence can be defined by utilizing different measures of
interference.  The weakest notion is just the consistency of the
probability sum rules that was called ``consistency'' by Griffiths
\cite{Gri84} and Omn\`es \cite{Omn92} and that term is used by some
to refer to all measures of interference. Vanishing of the real part of
\eqref{foursix} is a sufficient condition for the consistency of the
probability sum rules called the ``weak decoherence condition''.  We are
using the stronger condition \eqref{foursix} because it characterizes
widespread and typical mechanisms of decoherence.  Eq \eqref{foursix} has
been called the ``medium decoherence condition''.  ``Decoherence'' in
the context of this paper, thus, means the medium decoherence of sets of
histories.
In the literature the term ``decoherence'' 
has also been used to refer
to the decay in time of the off-diagonal elements of a reduced density
matrix defined by tracing the full density matrix over a given set of
variables \cite{Zur91}. The two notions of ``decoherence of reduced
density matrices'' and ``decoherence of histories'' are not generally
equivalent but also not unconnected in the sense
that in particular models
 certain physical processes can ensure both.  (See, \eg the remarks
in Section II.6.4 of \cite{Har91a}).}
  Probabilities can be
assigned to the individual histories in a decoherent set of alternative
histories because decoherence implies the probability sum rules
necessary for a consistent assignment.  The probability of an individual
history $\alpha$ is
\begin{equation}
p(\alpha) = \left\Vert C_\alpha |\Psi\rangle\right\Vert^2\, . 
\label{fourseven}
\end{equation}

To see how decoherence implies the probability sum rules, let us
consider an example in which there are just three sets of alternatives
at times $t_1, t_2$, and $t_3$.  A typical sum rule might be
\begin{equation}
\sum\nolimits_{\alpha_2} p\left(\alpha_3, \alpha_2, \alpha_1\right) =
p\left(\alpha_3, \alpha_1\right)\, .
\label{foureight}
\end{equation}
We show \eqref{foursix} and \eqref{fourseven} imply \eqref{foureight}.  To do that
write out the left hand side of \eqref{foureight} using \eqref{fourseven} and
suppress the time labels for compactness.
\begin{equation}
\sum\nolimits_{\alpha_2} p\left(\alpha_3, \alpha_2, \alpha_1\right) =
\sum\nolimits_{\alpha_2} \left\langle \Psi| P^1_{\alpha_1}
P^2_{\alpha_2} P^3_{\alpha_3} P^3_{\alpha_3} P^2_{\alpha_2}
P^1_{\alpha_1} |\Psi\right\rangle \, .
\label{fournine}
\end{equation}
Decoherence means that the sum on the right hand side of \eqref{fournine} can
be written with negligible error as
\begin{multline}
\sum\nolimits_{\alpha_2} p\left(\alpha_3, \alpha_2, \alpha_1\right)
 \\ \approx \sum\nolimits_{\alpha^\prime_2\alpha_2} \left\langle \Psi |
P^1_{\alpha_1}  P^2_{\alpha^\prime_2} P^3_{\alpha_3} P^3_{\alpha_3}
P^2_{\alpha_2}
P^1_{\alpha_1} |\Psi\right\rangle\, . 
\label{fourten}
\end{multline}
the extra terms in the sum being vanishingly small.
But now, applying the first of \eqref{fourone} we see
\begin{equation}
\sum\nolimits_{\alpha_2} p\left(\alpha_3, \alpha_2, \alpha_1\right)
\approx \left\langle \Psi| P^1_{\alpha_1}
 P^3_{\alpha_3} P^3_{\alpha_3} 
P^1_{\alpha_1} |\Psi\right\rangle = p\left(\alpha_3, \alpha_1\right)
\label{foureleven}
\end{equation}
so that the sum rule \eqref{foureight} is satisfied.

Given an initial state $|\Psi\rangle$ and a Hamiltonian $H$, one could,
in principle, identify all possible sets of decohering histories.  Among
these will be the exactly decohering sets where the orthogonality of the
branches is exact.  Indeed, trivial examples can be supplied by
resolving $|\Psi\rangle$ into a sum of orthogonal vectors at time $t_1$,
resolving those vectors into sums of further vectors such that the whole
set is orthogonal at time $t_2$, and so on.  However, such sets of
exactly decohering histories will not, in general, have a simple
description in terms of fundamental fields nor any connection, for
example, with the quasiclassical realm of familiar experience.  For
this reason sets of histories that approximately decohere are of
interest.  As we will argue in the next two Sections, realistic
mechanisms lead to the decoherence of histories constituting a
quasiclassical realm to an excellent approximation.  When the
decoherence condition \eqref{foursix} is approximately enforced, the
probability sum rules such as \eqref{foureight} will only be approximately
obeyed.  However, as discussed earlier, these probabilities for single
systems are meaningful up to the standard they are used.
  Approximate probabilities for which
the sum rules are satisfied to a comparable standard may therefore also
be employed in the process of prediction.  When we speak of approximate
decoherence and approximate probabilities we mean decoherence achieved
and probability sum rules satisfied beyond any standard that might be
conceivably contemplated for the accuracy of prediction and the
comparison of theory with experiment.

Decoherent sets of histories of the universe
are what we may utilize in the process of prediction
in quantum mechanics, for they
may be assigned probabilities.  Decoherence thus generalizes
and replaces the notion of ``measurement'', which served this role in
the Copenhagen interpretations.  Decoherence is a more precise, more
objective,
more observer-independent idea and gives a definite meaning to Everett's
branches.  For example, if their associated
histories
decohere, we may assign
probabilities to various values of reasonable scale density fluctuations
in the early universe whether or not anything like a ``measurement'' was
carried out on them and certainly whether or not there was an
``observer''
to do it.

\section{The Origins of Decoherence in Our Universe}

What are the features of coarse-grained sets of histories that
decohere in our universe?
In seeking to answer this question it is important to keep in mind the
basic
aspects of the theoretical framework on which decoherence depends.
Decoherence of a set of alternative histories is not a property of their
operators {\it alone}.  It depends  on the relations of those
operators to the initial state $|\Psi\rangle$, the Hamiltonian $H$, and the
fundamental
fields.  Given these, we could, in principle, {\it compute} which sets
of alternative histories decohere.

\begin{figure}[t]
\begin{center}
\epsfig{file=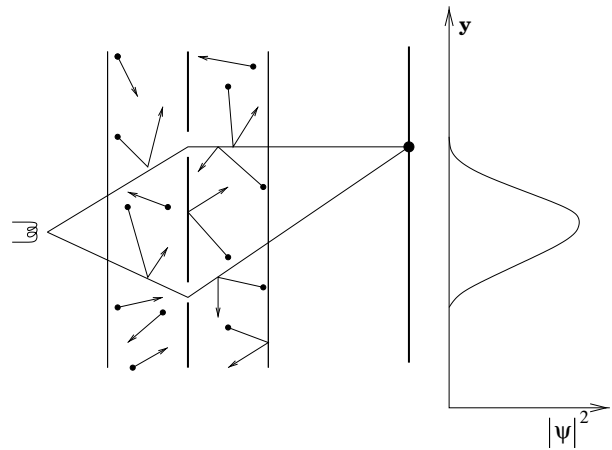,width=8cm,clip}
\caption{The two slit experiment
with an
interacting gas. Near the slits light particles of a gas collide with
the electrons.  Even if the collisions do not affect the trajectories of the
electrons very much they can still carry away the phase correlations
between
the histories in which the electron arrived at point $y$ on the screen
by passing through the upper slit and that in which it arrived at the same
point
by passing through the lower slit.  A coarse graining that described
only
of these two alternative histories of the electron would approximately
decohere as a consequence of the interactions with the gas given
adequate
density, cross-section, etc.  Interference is destroyed and
probabilities can be
assigned to these alternative histories of the electron in a way that
they
could not be if the gas were not present ({\it cf.} Fig. 1).  The lost phase
information is still available in correlations between states of the gas
and states of the electron.  The alternative histories of the electron would
not decohere in a coarse graining that included both the histories of the
electron
{\it and} operators that were sensitive to the correlations between the
electrons and the gas.\\
This model illustrates a widely occuring mechanism by which certain
types of
coarse-grained sets of alternative histories decohere in our universe.}
\end{center}
\end{figure}

We are not likely to carry out a computation of all decohering sets of
alternative histories for the universe, described in terms of the
fundamental
fields, anytime in the near future, if ever. It is therefore important
to investigate specific mechanisms by which decoherence occurs.
Let us begin with a very simple model due, in its essential features
 to Joos and Zeh \cite{JZ85}.
We consider the two-slit example again, but this time suppose that in the
neighborhood
of the slits there is a gas of photons or other light particles
colliding with
the electrons (Figure 4).  Physically it is easy to see what happens, the
random uncorrelated collisions can carry away delicate phase
correlations
between the beams even if the trajectories of the
electrons are not affected much.
  The interference pattern will then be destroyed and it will
be possible
to assign probabilities to whether the electron went through the upper
slit or
the lower slit.  

Let us see how this picture in words is given precise
meaning 
in  mathematics.  Initially, suppose the state of the entire
system is a state of the electron $|\psi >$ and $N$ distinguishable
``photons''
in states $|\varphi_1 \rangle$, $|\varphi_2 \rangle$, etc., {\it viz.~}
\begin{equation}
|\Psi \rangle = |\psi \rangle  |\varphi_1 \rangle |\varphi_2 >\cdots 
|\varphi_N\rangle
\, .
\label{fiveone}
\end{equation}
Suppose further that
$|\psi \rangle$ is a coherent superposition of a state in which the electron
passes
through the upper slit $|U \rangle$ and the lower slit $|L \rangle$. 
Explicitly:
\begin{equation}
|\psi \rangle = \alpha |U \rangle + \beta |L \rangle\, .
\label{fivetwo}
\end{equation}
Both states are wave packets in $x$, so that position in $x$
recapitulates history in time.  We now ask whether the history where the
electron passes through the upper slit and arrives at a detector at
point $y$ on the screen, decoheres from that in which it passes through
the lower slit and arrives at point $y$ as a consequence of the initial
condition of this ``universe''.  That is, as in Section 4, we ask
whether the two branches
\begin{equation}
P_y(t_2) P_U(t_1) |\Psi\rangle\quad , \quad P_y(t_2) P_L(t_1)
 |\Psi\rangle 
\label{fivethree}
\end{equation}
are nearly orthogonal, the times of the projections being those for the
nearly classical motion in $x$.  We work this out in the
Schr\"odinger picture where the initial state evolves, and the
projections on the electron's position are applied to it at the
appropriate times.

Collisions occur, but the states $|U\rangle$
 and  $|L \rangle$ are left more
or less undisturbed.  The states of the ``photons'', of course, are
significantly affected.  If the photons are dilute enough to be
scattered once
by the electron in its time to traverse the gas the two branches \eqref{fivethree}
will be
approximately
\begin{subequations}
\label{fivefour}
\begin{equation}
\alpha P_y |U \rangle S_U |\varphi_1 \rangle S_U |\varphi_2 \rangle
\cdots S_U |\varphi_N\rangle
\, , 
\label{fivefoura}
\end{equation}
and
\begin{equation}
\beta\ P_y |L \rangle S_L |\varphi_1 \rangle S_L |\varphi_2 \rangle
\cdots S_L |\varphi_N\rangle
\, .
\label{fivefourb}
\end{equation}
\end{subequations}
Here, $S_U$ and $S_L$ are the scattering matrices from an electron in
the
vicinity of the upper slit and the lower slit respectively.  The two
branches
 in \eqref{fivefour} decohere because the states of the ``photons'' are nearly
orthogonal.
The overlap of the branches is proportional to
\begin{equation}
\langle\varphi_1 |S^{\dagger}_U S_L |\varphi_1 \rangle\langle
\varphi_2 | S^{\dagger}_U
S_L |\varphi_2
\rangle\cdots \langle\varphi_N |S^\dagger_U S_L\ |\varphi_N \rangle
\, .
\label{fivefive}
\end{equation}
Now, the $S$-matrices for scattering off the upper position or the lower
position
can be connected to that of an electron at the orgin by a translation
\begin{subequations}
\label{fivesix}
\begin{eqnarray}
S_U & = & \exp(-i{\bf k} \cdot{\bf x}_U) S\ \exp(+i {\bf k} \cdot  {\bf
x}_U)\, ,
\label{fivesixa}\\
S_L & = & \exp(-i{\bf k} \cdot{\bf x}_L) S\ \exp(+i {\bf k} \cdot  {\bf
x}_L)\ .
\label{fivesixb}
\end{eqnarray}
\end{subequations}
Here, $\hbar{\bf k}$ is the momentum of a photon, ${\bf x}_U$ and ${\bf
x}_L$
are the positions of the slits and $S$ is the  scattering matrix from
an electron at the origin.
\begin{equation}
\langle{\bf k}^\prime |S|{\bf k}\rangle = \delta^{(3)}\bigl({\bf k} - {\bf
k}^\prime
\bigr)
+ \frac{i}{2\pi\omega_{\bf k}} f\bigl({\bf k}, {\bf k}^\prime\bigr)
\delta
\bigl(\omega_k - \omega^\prime_k\bigr)\, ,
\label{fiveseven}
\end{equation}
where $f$ is the scattering amplitude and $\omega_k = |\vec k|$.

Consider the case where initially
 all the photons are in plane wave states in an
interaction volume $V$, all having the same energy $\hbar\omega$, but
with
random
orientations for their momenta.  Suppose further that the energy is low
so that
the electron is not much disturbed by a scattering
and low enough so the wavelength is
much longer than the separation between the slits, $k|{\bf x}_U - {\bf
x}_L | \ll 1$.
 It is then possible to work out the overlap.  The answer
according to
Joos and Zeh \cite{JZ85} is
\begin{equation}
\left(1-\frac{(k|{\bf x}_U - {\bf
x}_L|)^2}{8\pi^2V^{2/3}}\sigma\right)^N
\label{fiveeight}
\end{equation}
where $\sigma$ is the effective scattering cross section and the
individual terms have been averaged over incoming directions.  Even if
$\sigma$ is small, as $N$ becomes large this tends to zero.  
In this way decoherence becomes a
quantitative phenomenon.

What such models convincingly show is that decoherence is frequent and
widespread in the universe for histories of certain kinds of variables.  
Joos and Zeh  calculate that a
superposition of two positions of a grain of dust, 1mm apart, is
decohered simply by the scattering of the cosmic background radiation on
the timescale of a nanosecond.  The existence of such mechanisms means
that the only realistic isolated systems are of cosmological dimensions.
So widespread is this kind of phenomena with the
initial condition and dynamics of our universe, that we may meaningfully
speak of habitually decohering variables such as the center of mass
positions of massive bodies.

\section{The Copenhagen Approximation}

What is the relation of the familiar Copenhagen quantum mechanics
described in Section III to the more general ``post-Everett'' quantum
mechanics of closed systems described in Sections IV and V?
Copenhagen quantum mechanics predicts the probabilities of the histories
of measured subsystems.  Measurement situations may be described in
a closed system that contains both measured
subsystem and measuring apparatus.  In a typical measurement situation
the values of a variable not normally decohering become correlated with
alternatives of the apparatus that decohere because of {\it its}
 interactions
with the rest of the closed system.  The correlation means
that the measured alternatives decohere because the alternatives of the
apparatus with which they are correlated  decohere. 

The recovery of the Copenhagen rule for when probabilities may be
assigned is immediate.  Measured quantities are correlated with
decohering histories.  Decohering histories can be assigned
probabilities.  Thus in the two-slit experiment (Figure 1), when the
electron interacts with an apparatus that determines which slit it
passed through, it is the decoherence of the alternative configurations
of the apparatus that enables probabilities to be assigned for the
electron. 

There is nothing incorrect about Copenhagen quantum
mechanics.  Neither is it, in any sense, opposed to the post-Everett
formulation of the quantum mechanics of closed systems.
 It is an {\it approximation} to the more general framework
appropriate in the special cases of measurement situations and when the
decoherence of alternative configurations of the apparatus  may be
idealized as exact and instantaneous.  However, while measurement
situations imply decoherence, they are only special cases of decohering
histories.  Probabilities may be assigned to alternative positions of the
moon and to  alternative values of
density fluctuations near the big bang in a universe in
which these alternatives decohere, whether or not they were participants in a
measurement situation and certainly whether or not there was an observer
registering their values.

\section{Quasiclassical Realms}

As observers of the universe, we deal with coarse-grained histories that
reflect our
own limited sensory perceptions, extended by instruments, communication
and
records  but in the end characterized by a large amount of ignorance.
Yet, we have the impression that the universe exhibits a much finer-grained
set of histories,
independent of us,
defining an always decohering ``quasiclassical realm'',
to which our senses are adapted, but deal with only a small part of it.
If we are preparing for a journey into a yet unseen part of the
universe, we do not believe that we need to equip ourselves with
spacesuits having detectors sensitive, say, to coherent superpositions of
position or other unfamiliar quantum variables.  We expect that the
familiar quasiclassical variables will decohere and be approximately
correlated in time by classical deterministic laws in any new part of
the universe we may visit just as they are here and now.

Since the post-Everett quantum mechanics of closed systems does not
posit a quasiclassical realm, it must provide an explanation of this
manifest fact of everyday experience.  No such explanation can be
provided from the dynamics of quantum theory alone.  Rather, like
decoherence, the existence of a quasiclassical realm in the universe
must be a consequence of both initial
condition of the universe and the Hamiltonain describing evolution.

Roughly speaking, a quasiclassical realm should be a set of alternative
histories that decoheres according to a realistic principle
of decoherence, that is maximally refined  consistent with that notion of
 decoherence, and
whose individual histories are described largely by alternative values
of a limited set of quasiclassical variables at different moments of
time that exhibit as much as possible patterns of
classical
correlation in time.
To make the question of the existence of one or more quasiclassical
realms into a {\it calculable} question in quantum cosmology 
 we need measures of how close a set of histories comes to
constituting a ``quasiclassical realm''.  A quasiclassical realm 
cannot be a
    {\it completely}
fine-grained description for then it would not decohere.  It cannot
consist
{\it entirely} of a few ``quasiclassical variables'' repeated over and over
because
sometimes
we may measure something highly quantum mechanical.  Quasiclassical
variables
cannot be
{\it always} correlated in time by classical laws because sometimes
quantum
mechanical phenomena cause deviations from classical physics.  We need
measures
for maximality and classicality \cite{GH90a}.

It
is possible to give crude arguments for the type of
habitually decohering
operators we expect to occur over and over again in a set of
histories defining a quasiclassical realm \cite{GH90a}. 
 Such habitually decohering
operators are called ``quasiclassical operators''.
In the earliest instants of the universe the operators defining
spacetime on
scales well above the Planck scale emerge from the quantum fog as
quasiclassical.  Any theory of the initial condition that does
not
imply this is simply inconsistent with observation in a manifest way.
A background spacetime is thus defined and conservation laws arising
from its symmetries have meaning. Then,
where there are suitable conditions of low temperature, density,
etc., various sorts of
hydrodynamic variables may emerge as quasiclassical operators.  These
are
integrals over suitably small volumes of densities of conserved or
nearly
conserved quantities.  Examples are densities of energy, momentum,
baryon
number, and, in later epochs, nuclei, and even chemical species.  The
sizes of
the volumes are limited above by maximality and are limited below by
classicality
because they require sufficient ``inertia'' resulting from their
approximate conservation  to enable them to resist
deviations
from predictability caused by their interactions with one another, by
quantum
spreading, and by the quantum and statistical fluctuations resulting
from
interactions with the rest of the universe that accomplish decoherence
\cite{GH92}.
Suitable integrals of densities of
approximately conserved quantities are thus candidates for habitually
decohering quasiclassical operators.
These ``hydrodynamic variables'' {\it are} among the principal variables
of classical physics.

It would be in such ways that the quasiclassical realm of familiar
experience would be an emergent property of the fundamental description of 
the universe, not generally in quantum mechanics,
but as a consequence of our specific initial condition and the
Hamiltonian
describing evolution.  Whether the universe exhibits a quasiclassical
realm, and, indeed, whether it exhibits more than one essentially
inequivalent realm, thus become calculable questions in the quantum
mechanics of closed systems.

\section{Conclusion}

Quantum mechanics is best and most fundamentally understood in
the context of quantum mechanics of closed systems, most generally the
universe as a whole.  The founders of quantum mechanics
were right in pointing out that something external to the framework of
wave function and the Schr\"odinger equation {\it is} needed to
interpret the theory.  But it is not a postulated classical realm to
which quantum mechanics does not apply.  Rather it is the initial
condition of the universe that, together with the action function of the
elementary particles and the throws of the quantum dice since the
beginning, is the likely origin of quasiclassical realm(s) within
quantum theory itself.

\acknowledgments

The formulation of quantum mechanics described in this paper is a result
of joint work with Murray Gell-Mann.  It is a pleasure to thank him for
the many conversations over the years and for permission to summarize
aspects of our work here. Preparation of the report was supported in
part by the National Science Foundation under grant PHY90-08502.

\end{document}